\documentclass[prd,preprint,superscriptaddress,amsmath,amssymb,nofootinbib]{revtex4}
\usepackage{graphicx}
\usepackage{dcolumn}
\usepackage{bm}
\usepackage{amssymb}
\usepackage{amsmath}
\usepackage{epsfig}    
\usepackage{color}
\usepackage{slashed}
\usepackage{hhline}

\def\be{\begin{equation}}
\def\ee{\end{equation}}
\newcommand{\bea}{\begin{eqnarray}}
\newcommand{\eea}{\end{eqnarray}}
\newcommand{\nn}{\nonumber}

\begin{document}

 \begin{flushright} {KIAS-P18084}, APCTP Pre2018 - 012  \end{flushright}

\title{One-loop neutrino mass model without any additional symmetries}

\author{Takaaki Nomura}
\email{nomura@kias.re.kr}
\affiliation{School of Physics, KIAS, Seoul 02455, Republic of Korea}

\author{Hiroshi Okada}
\email{hiroshi.okada@apctp.org}
\affiliation{Asia Pacific Center for Theoretical Physics (APCTP) - Headquarters San 31, Hyoja-dong,
Nam-gu, Pohang 790-784, Korea}

\date{\today}

\begin{abstract}
We propose a radiative seesaw model at one-loop level, introducing fields with large multiplet of $SU(2)_L$. Thanks to large representations, any additional symmetries are not needed. In this framework, we formulate lepton and new fermion sector such as mass matrices, LFVs, and muon $g-2$. 
Furthermore, we show our cut-off scale via RGEs of $SU(2)_L$, and numerical analysis at a benchmark point included in dark matter candidate.  
Finally, we briefly discuss processes producing exotic particles in our model via proton-proton collision and possibility of detecting them at the large hadron collider experiments.
 \end{abstract}
\maketitle

\section{Introductions}
Radiative seesaw models are promising candidates to connect active neutrinos and dark matter (DM) candidate in addition to the natural explanation of smallness for neutrino masses.
In order to realize such kinds of models, typically additional symmetries such as $Z_N$ symmetry are simply needed for stabilizing the DM candidate as well as forbidding neutrino masses at tree level~\cite{Ma:2006km, Krauss:2002px, Aoki:2008av, Gustafsson:2012vj, Kajiyama:2013zla}.
As an alternative approach, unique charge assignments under $SU(2)_L$~{\cite{Nomura:2017abu,Nomura:2016dnf, Nomura:2016jnl,Anamiati:2018cuq}} and/or $U(1)_Y$~\cite{Zee, Li, Zee-Babu, Cheung:2016fjo, Nomura:2016ask, Cheung:2018itc, Cheung:2017kxb} in the standard model (SM) are applied to new fields for restricting interactions, without additional symmetries.
In particular, introduction of large $SU(2)_L$ multiplets is interesting since their interactions are strongly restricted by gauge symmetry and rich phenomenology is provided by such new particles. 

In this paper, we introduce several exotic fields with large representations of $SU(2)_L$ as shown in tab.~\ref{tab:1}.
Then we find that these field-contents lead to a reasonable radiative seesaw model due to assuring stability of the inert quintet scalar field $H_5$ in a renormalizable theory without any additional symmetries, where {this issue has been discussed by the comprehensive paper~\cite{Cirelli:2005uq}.}
It is also found that cutoff scale of a theory would be much lower than the Planck scale by analyzing running of gauge coupling with renormalization groups since 
contributions from fields with large $SU(2)_L$ representations are sizable.
This feature would be dangerous to obtain reliable predictions as we would have several unknown resonances around  cutoff  scale other than new particles introduced in our model. 
Then we find that new fields with large $SU(2)_L$ representations should be sufficiently heavy; 
for example, if its scale is 5 TeV the cutoff scale is larger than $10^3$ TeV and our predictions are reliable at the LHC experiments.
On the other hand we might be able to test our model by testing deviation of the SM prediction for running of gauge coupling at around TeV or higher scale.

This letter is organized as follows.
In Sec. II, {we review our model and formulate the lepton and new fermion sector. 
In Sec. III we show numerical analysis at a benchmark point and discuss collider physics.
Finally we devote the summary of our results and the conclusion.}

\section{Model setup and Constraints}
\begin{table}[t!]
\begin{tabular}{|c||c|c|c|c||c|c|}\hline\hline  
& ~$L_L^a$~& ~$e_R^a$~& ~$\psi^a$~& ~$\Sigma_R^a$~& ~$H_2$~& ~$H_5$~\\\hline\hline 
$SU(2)_L$   & $\bm{2}$  & $\bm{1}$  & $\bm{4}$  & $\bm{5}$ & $\bm{2}$  & $\bm{5}$   \\\hline 
$U(1)_Y$    & $-\frac12$  & $-1$ & {-$\frac12$}  & $0$  & $\frac12$   &{$0$} \\\hline
\end{tabular}
\caption{Charge assignments of the our lepton and scalar fields
under $SU(2)_L\times U(1)_Y$, where the upper index $a$ is the number of family that runs over 1-3 and
all of them are singlet under $SU(3)_C$. }\label{tab:1}
\end{table}

In this section we formulate our model.
As for the fermion sector, we introduce three families of vector fermions $\psi$ with $(4,{-1/2})$ charge under the $SU(2)_L\times U(1)_Y$ gauge symmetry, and right-handed fermions $\Sigma_R$ with $(5,0)$ charge under the $SU(2)_L\times U(1)_Y$ gauge symmetry.
As for the scalar sector, we add a quintet complex scalar field $H_5$ with {$0$} charge under the $U(1)_Y$ gauge symmetry, where SM-like Higgs field is denoted as $H_2$.
Here we write vacuum expectation value{(VEV)} of $H_2$ by  {$\langle H_2\rangle\equiv (0, v/\sqrt2)^T$}
which induces the spontaneously electroweak symmetry breaking.
All the field contents and their assignments are summarized in Table~\ref{tab:1}, where the quark sector is exactly the same as the SM.
The renormalizable Yukawa Lagrangian under these symmetries is given by
\begin{align}
-{\cal L_\ell}
& =  y_{\ell_{aa}} \bar L^a_L H_2 e^a_R  +  f_{ab} [ \bar L^a_L H_5 (\psi_R)^b ]  
 +  g_{ab} [\bar \psi^{a}_L  H_2^* \Sigma^b_R]
+ g'_{ab} [(\bar\psi^c_R)^a  H_2\Sigma^b_R] \nn\\
&+M_{D_{aa}} \bar \psi^a_R \psi_L^a+\frac12 M_{\Sigma_{aa}}(\bar \Sigma^c_R)^a \Sigma_R^{a}
+ {\rm h.c.}, \label{Eq:yuk}
\end{align}
where $SU(2)_L$ index is omitted assuming it is contracted to be gauge invariant inside bracket [$\cdots$], 
upper indices $(a,b)=1$-$3$ are the number of families, and ($y_\ell,M_D, M_\Sigma$) are assumed to be  diagonal matrix with real parameters without loss of generality.
The mass matrix of charged-lepton is defined by $m_\ell=y_\ell v/\sqrt2$. 
Here we assign lepton number $1$ to $\psi_{L(R)}$ and $0$ to $\Sigma_R$ so that the source of lepton number violation is terms with coupling $g_{ab}$ and $g'_{ab}$ in the Lagrangian.

\noindent \underline{\it Scalar potential and VEVs}:
The scalar potential in our model is given by
\begin{align}
{\cal V} = & -\mu_h^2 |H_2|^2  + \lambda_H |H_2|^4 + M_5^2 |H_5|^2+ M'^2_5 [H_5]^2  
+\mu[H_5^3+{\rm h.c.}]
\nn\\
& + \sum_i \lambda_0^i [H_5 H_5 H_5 H_5 + h.c.]
 + \lambda_{H}|H_2|^4 + \lambda_{H_5}|H_5|^4  + \lambda_{H_2H_5}|H_2|^2|H_5|^2, 
\label{Eq:potential}
\end{align}
where sum in quartic term of $H_5$ indicate independent terms corresponding to different contraction of $SU(2)_L$ index.
\footnote{Due to the trilinear term of $\mu$ in the scalar potential, the DM candidates for $\psi, \Sigma, H_5$ do not have any remnant symmetries after the electroweak symmetry breaking. Nevertheless, they could be a DM candidate that is called "minimal dark matter"~\cite{Cirelli:2007xd}. }
Applying condition $\partial {\cal V}/\partial v = 0$, we obtain the VEV of Higgs field $H_2$ as 
\begin{align}
v \sim \sqrt{\frac{\mu^2_h}{\lambda_H}}.
\end{align}
{On the other hand, we require a non-zero VEV of $H_5$ and stability of the potential by imposing $M_{5}^2 >0$, $\mu >0$ and $\{\lambda_0^i, \lambda_{H_5} \} > 0$ for parameters in the potential. }

\noindent \underline{\it Exotic particles} :
The scalars and fermions with large $SU(2)_L$ multiplet provide exotic charged particles.
Here we write components of multiplets as
\begin{align}
& H_5 = (\phi_5^{++}, \phi_5^{+}, \phi_5^{0}, \phi'^{-}_5, \phi'^{--}_5)^T,  \label{eq:H5} \\
& \psi_{L(R)} = (\psi^{+}, \psi^{0}, \psi'^{-}, \psi^{--})^T_{L(R)},  \label{eq:psiLR} \\
& \Sigma_R = (\Sigma^{++}, \Sigma^{+}, \Sigma^{0}, \Sigma'^{-}, \Sigma'^{--})_R^T. \label{eq:sigmaR} 
\end{align}
The mass of component in $H_5$ is given by $\sim M_5$ where charged particles in the same multiplet have degenerate mass at tree level which will be shifted at loop level~\cite{Cirelli:2005uq}. 
For charged fermions, components from $\psi_{L(R)}$ and $\Sigma_R$ can be mixed after electroweak symmetry breaking via Yukawa coupling.
If the Yukawa couplings are negligibly small  
the charged components in $\psi_{L(R)}$ have Dirac mass $M_D$ while
the charged components in $\Sigma_R$ have Dirac mass $M_\Sigma$ where mass terms are constructed by pairs of positive-negative charged components in the multiplet.
Note that mass term of neutral component is discussed with neutrino sector below.

\begin{figure}[tb]
\begin{center}
\includegraphics[width=10.0cm]{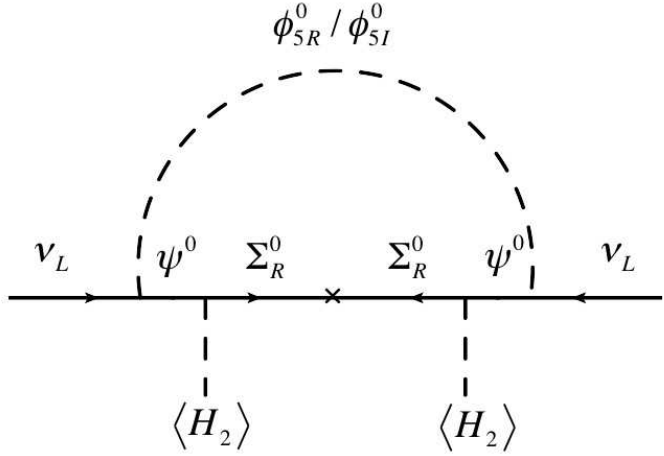}
\caption{Feynman diagram to generate active neutrino mass.}
\label{fig:diagram}
\end{center}\end{figure}

\subsection{Neutral fermion masses}
\noindent \underline{\it Heavier neutral sector}:
After the spontaneously symmetry breaking, neutral fermion mass matrix in basis of $(\Psi^0_R\equiv (\psi^0_R,\psi_L^{0c},\Sigma_R^0)^T$ is given by
\begin{align}
M_N
&=
\left[\begin{array}{ccc}
0 & M_D^T & m'  \\ 
M_D & 0 & m \\ 
m'^T  & m^T & \frac12(M_\Sigma+ M^\dag_{\Sigma}) \\ 
\end{array}\right],
\end{align}
where $m^{(')}\equiv g^{(')} v/\sqrt2$.
Then $M_N$ are $\Psi^0_R$ are respectively rotated by the unitary matrix as
\begin{align}
\Psi^0_R=V^T (\psi_{1-9})_R,\quad D_N\equiv{\rm diag}(M_1,...,M_9)=V M_N V^T,
\end{align}
where $(\psi_{1-9})_R$ and $D_N$ are respectively mass eigenvectors and mass eigenvalues.
We thus have 9 exotic Majorana fermions with mass eigenvalue $M_{1-9}$.

\noindent \underline{\it  Active neutrino sector} :
 In our scenario, we assume lepton number conservation is violated only through Yukawa interactions with coupling $g$ and $g'$ in Eq.~(\ref{Eq:yuk}), 
and physics at scale higher than cut-off does not contribute lepton number violation. 
In such a case, active neutrino mass can only be induced via interactions in Eq.~(\ref{Eq:yuk}) and (\ref{Eq:potential}) even if cut-off scale is not very high as we discuss below.
Then active neutrino mass is dominantly generated at one-loop level where $\psi_\alpha$ and $H_5$ propagate inside a loop diagram as shown Fig.~\ref{fig:diagram}.
As a result the active neutrino mass matrix is obtained such that
\begin{align}
m_\nu = \sum_{\alpha=1-9} \frac{F_{i\alpha} F^T_{\alpha j} M_\alpha}{32\pi^2}
\left[
\frac{m^2_R}{m^2_R-M^2_\alpha}\ln\left[\frac{m^2_R}{M^2_\alpha}\right]
-
\frac{m^2_I}{m^2_I-M^2_\alpha}\ln\left[\frac{m^2_I}{M^2_\alpha}\right]
\right],
\end{align}
where $F_{i\alpha} \equiv \sum_{k=1}^3 f_{ik} V^T_{k\alpha}$ and $m_{R(I)}$ is mass of $\phi^0_{5R(I)}$ which comes from real/imaginary components of $\phi_{5}^0$. 
Diagonalizing the matrix, neutrino mass eigenvalues ($D_\nu$) are found as $D_\nu=U_{\rm MNS} m_\nu U^T_{\rm MNS}$, where $U_{\rm MNS}$ is  the MNS matrix. Once we define $m_{\nu} \equiv F R F^T$, one can rewrite $f$ in terms of the other parameters~\cite{Casas:2001sr, Chiang:2017tai} as follows:
\begin{align}
f_{ik}=\sum_{\alpha=1}^9 U^\dag_{ij} \sqrt{D_{\nu_{jj}}} O_{j\alpha} \sqrt{R_{\alpha\alpha}} V^*_{\alpha k},
\end{align}
where $O$ is a three by nine arbitrary matrix, satisfying $OO^T=1$. 

\subsection{Charged fermion masses}
\noindent \underline{\it  Singly-charged fermion sector}:
The singly-charged fermion mass matrix, in basis of $\Psi^-_R \equiv (\psi^{-}_R (\equiv (\psi_L^{+})^c ),\psi'^-_R,\Sigma'^-_R)^T$ and $\Psi^-_L \equiv (\psi^{-}_L ,\psi'^-_L,\Sigma^-_L (\equiv (\Sigma_R^+)^c ))^T$, is given by
\begin{align}
L_{M_\pm} = \bar \Psi^-_L M_\pm \Psi^-_R, \quad
M_\pm
&=
\left[\begin{array}{ccc}
M_D^T & 0 & -\frac12 m'  \\ 
0 & M_D & \frac{\sqrt3}{2}m \\
-\frac12 m'^T  & \frac{\sqrt3}2 m^T & \frac12(M_\Sigma+ M^T_{\Sigma})\\ 
\end{array}\right].
\end{align}
When $M_\pm$ is symmetric,  
$M_\pm$ are $\Psi^\pm_{L(R)}$ and respectively rotated by the unitary matrix as
\begin{align}
\Psi^\pm_{L(R)}=V^T_C \psi^\pm_{{L(R)}_{1-9}},\quad D_\pm\equiv{\rm diag}(M_{C_1},...,M_{C_{9}})=V_C M_\pm V_C^T,
\end{align}
where $\psi^\pm_{R_{1-9}}$ and $D_\pm$ are respectively mass eigenvectors and mass eigenvalues of Dirac type.


\noindent \underline{\it  Doubly-charged fermion sector}:
The doubly-charged fermion mass matrix, in basis of $\Psi^{--}_R \equiv ( \psi_R^{--},\Sigma'^{--}_R)^T$ and $\Psi^{--}_L \equiv ( \psi_L^{--} ,\Sigma^{--}_L (\equiv (\Sigma_R^{++})^c ))^T$, is given by
\begin{align}
L_{M_{\pm \pm}} = \bar \Psi_L^{--} M_{\pm \pm} \Psi^{--}_R, \quad
M_{\pm\pm} =
\left[\begin{array}{cc}
M_D  &  m  \\ 
m'^T & \frac12 (M_\Sigma + M^T_\Sigma)  \\
\end{array}\right].
\end{align}
When $M_{\pm\pm}$ is symmetric,  
then $M_{\pm\pm}$ and $\Psi^{\pm\pm}_{L(R)}$ are respectively rotated by the unitary matrix as
\begin{align}
\Psi^{\pm\pm}_{L(R)}=V^T_{CC} \psi_{{L(R)}_{1-6}}^{\pm\pm},\quad D_{\pm \pm} \equiv{\rm diag}(M_{CC_1},...,M_{CC_{6}})=V_{CC} M_{\pm\pm} V_{CC}^T,
\end{align}
where $ \psi_{{L(R)}_{1-6}}^{\pm\pm}$ and $D_{\pm\pm}$ are respectively mass eigenvectors and mass eigenvalues of Dirac type.


 \subsection{Constraints from running of gauge coupling and LFV}
 
\begin{figure}[tb]
\begin{center}
\includegraphics[width=10.0cm]{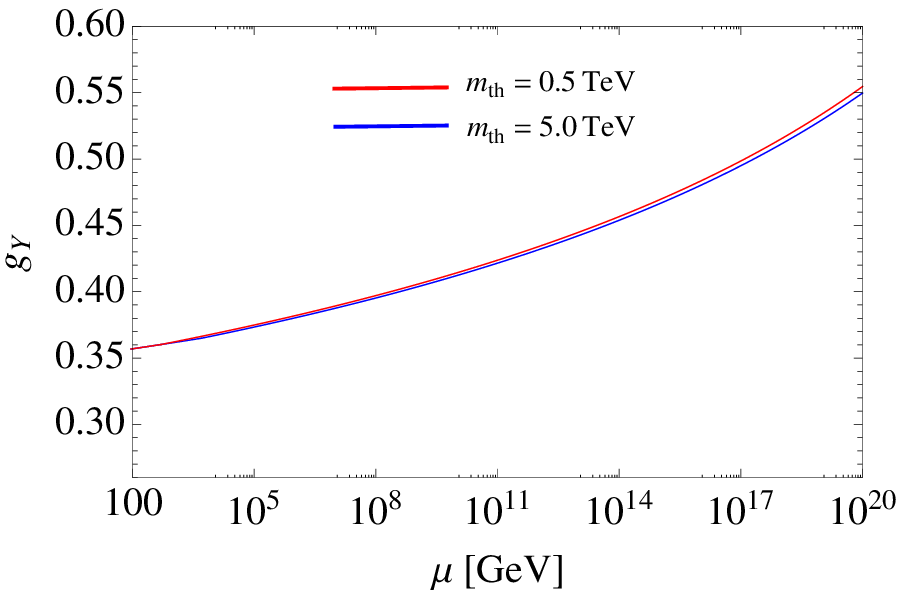}
\includegraphics[width=10.0cm]{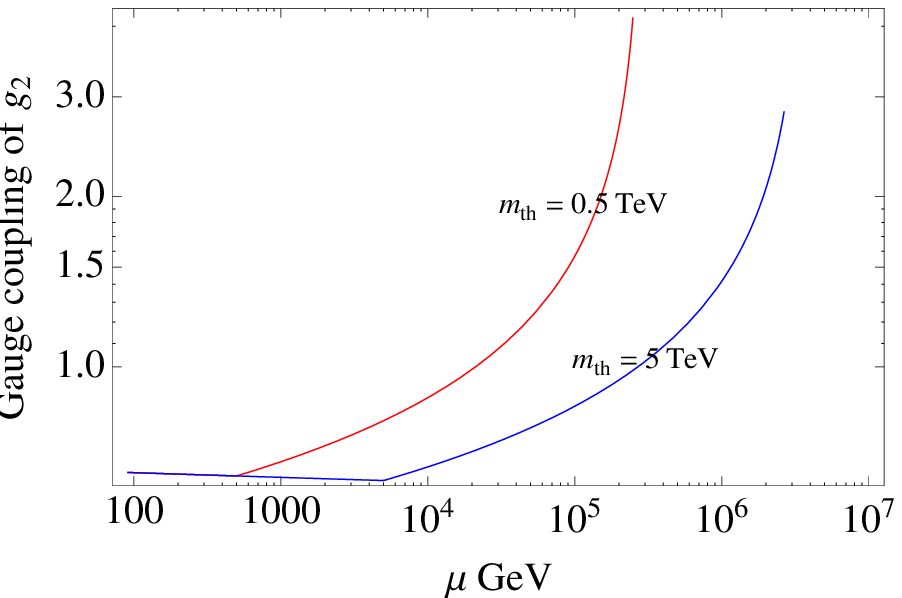}
\caption{The running of $g_Y$ and $g_2$ in terms of a reference energy of $\mu$, where the red line corresponds to  $m_{th}=$0.5 TeV,
while the blue one does  $m_{th}=$5 TeV. }
\label{fig:rge}
\end{center}\end{figure}

 {
\noindent \underline{\it Beta function of $SU(2)_L$ and $U(1)_Y$ gauge coupling $g_2$ and $g_Y$}
\label{beta-func}
Here it is worth  discussing the running of gauge couplings of $g_2$ and $g_Y$ in the presence of several new multiplet fields of $SU(2)_L$ and new $U(1)_Y$ charged fields. 
The new contribution to $g_2$ for an $SU(2)_L$ quintet fermion(boson) $\psi(H_4)$, septet fermion $\Sigma_R$, and quartet boson $H_5$ are respectively given by
\begin{align}
 \Delta b^{\psi}_{g_2}=\frac{10}{3}, \ \Delta b^{\Sigma_R}_{g_2}=\frac{20}{3},\ \Delta b^{H_5}_{g_2}=\frac{10}{3}  .
\end{align}
In addition, new contribution to $g_Y$ from $\psi_a$ is given by
\begin{equation}
 \Delta b^{\psi}_{g_Y}= 2
\end{equation}
Then one finds the energy evolution of the gauge coupling $g_2$ and $g_Y$ as~\cite{Nomura:2017abu, Kanemura:2015bli}
\begin{align}
&\frac{1}{g^2_{2}(\mu)}=\frac1{g_2^2(m_{in})}-\frac{b^{SM}_{g_2}}{(4\pi)^2}\ln\left[\frac{\mu^2}{m_{in}^2}\right]
-\theta(\mu-m_{th}) 
 \frac{(N_{f_\psi} \Delta b^{\psi}_{g_2}+ N_{f_\Sigma}\Delta b^{\Sigma_R}_{g_2}) +\Delta b^{H_5}_{g_2}}{(4\pi)^2}\ln\left[\frac{\mu^2}{m_{th}^2}\right], \nonumber \\
&\frac{1}{g^2_{Y}(\mu)}=\frac1{g_Y^2(m_{in})}-\frac{b^{SM}_{g_Y}}{(4\pi)^2}\ln\left[\frac{\mu^2}{m_{in}^2}\right]
-\theta(\mu-m_{th}) 
 \frac{N_{f_\psi} \Delta b^{\psi}_{g_Y} }{(4\pi)^2}\ln\left[\frac{\mu^2}{m_{th}^2}\right],
\label{eq:rge_g}
\end{align}
where $N_{f_{\psi/\Sigma}}=3$ is the number of $\psi$ and $\Sigma_R$, $\mu$ is a reference energy, $b^{SM}_{g_Y}=41/6$, $b^{SM}_{g_2}=-19/6$, and we assume to be $m_{in}(=m_Z) < m_{th}$, being $m_{th}$ threshold masses of exotic fermions and bosons.
The resulting flow of ${g_2}(\mu)$ and $g_Y(\mu)$ are then given by the Fig.~\ref{fig:rge}.
For $g_Y$, we find it is relevant up to Planck scale.
For $g_2$, the figure shows that the red line is relevant up to the mass scale $\mu={\cal O}(100)$ TeV in case of $m_{th}=$0.5 TeV,
while the blue is relevant up to the mass scale $\mu={\cal O}(1)$ PeV in case of $m_{th}=$5 TeV.
Thus our theory does not spoil, as far as we work on at around the scale of TeV; see also ref.~\cite{Sierra:2016rcz}. }

\noindent \underline{\it  Lepton flavor violations(LFVs)}
LFVs arise from the term $f$ at one-loop level, and its form can be given by~\cite{Lindner:2016bgg, Baek:2016kud}
\begin{align}
 {\rm BR}(\ell_i\to\ell_j\gamma)= \frac{48\pi^3\alpha_{\rm em} C_{ij} }{{\rm G_F^2} m_{\ell_i}^2}\left(|a_{R_{ij}}|^2+|a_{L_{ij}}|^2\right),
 \end{align}
where 
 \begin{align}
 a_{R_{ij}} &=\frac{m_{\ell_i}} {(4\pi)^2}
\left[\frac34 F_{j\alpha} F^\dag_{\alpha i} G(\psi_\alpha,H^-_5)
+F^{(\pm)}_{j\beta} F^{(\pm)\dag}_{\beta i} [ 2 G(\psi^\pm_\beta,H^{\pm\pm}_5) + G(H^{\pm\pm}_5,\psi^\pm_\beta)] 
\right.\nn\\
&- \left. \frac12 F'^{(\pm)}_{j\gamma} F'^{(\pm)\dag}_{\gamma i} G(H_5^0, \psi^\pm_\gamma)
-\frac14 F^{(\pm\pm)}_{j\rho} F^{(\pm\pm)\dag}_{\rho i} [G(\psi^{\pm\pm}_\rho,H^{\pm}_5) + 2 G(H^{\pm}_5,\psi^{\pm\pm}_\rho)] 
\right],
\label{eq:aR}
\end{align}
$a_L=a_R(m_{\ell_i}\to m_{\ell_j})$, $F^{(\pm)}_{i\beta}\equiv \sum_{j=1}^3f_{ij} (V^T_C)_{j\beta}$,
$F'^{(\pm)}_{i\gamma}\equiv \sum_{j=4}^6 f_{ij} (V^T_C)_{j\gamma}$, $F^{(\pm\pm)}_{i\rho}\equiv \sum_{j=1}^3f_{ij} (V^T_{CC})_{j\rho}$,
and 
\begin{align}
&G(a,b)\equiv \int_0^1dx\int_0^{1-x}dy\frac{xy}{(x^2-x)m^2_{\ell_i} +x m_a^2+(1-x) m^2_b}.
\end{align}
LFV with higher $SU(2)_L$ representation in radiative neutrino mass model are also discussed in ref.~\cite{Chowdhury:2015sla,Chowdhury:2018nhd}.

\noindent \underline{\it New contributions to the muon anomalous magnetic moment (muon $g-2$: $\Delta a_\mu$)} 
In our model $\Delta a_\mu$ arises from the same interactions inducing LFVs and can be formulated by the following expression:
\begin{align}
&\Delta a_\mu \approx -m_\mu [{a_{L_{\mu\mu}}+a_{R_{\mu\mu}}}] 
= -2m_\mu{a_{L_{\mu\mu}}}, \label{eq:G2-ZP}
\end{align}
where we use the fact that $a_{L_{\mu \mu}} = a_{R_{\mu \mu}}$. 
Thus first two terms in Eq.~(\ref{eq:aR}) provide negative contribution while the last two terms give positive contribution.
We will compare our prediction to the $1 \sigma$ range of current estimation, $\Delta a_\mu= (26.1\pm8.0)\times 10^{-10}$, in our numerical analysis below.

\subsection{perturbation of heavier fermions}
Before numerical analysis, let us introduce mass eigenvectors and eigenvalues with perturbation theory up to the first order.
In this case, one finds analytical formulations for complicated mass matrix.
First of all, we assume related mass matrices $m,m', M_D, M_\Sigma$ to be real and diagonal for simplicity.~\footnote{Although this simplification is ad hoc, any phenomenologies that we focus on would not be spoiled!}
{\it Moreover, we impose hierarchy among them; $m,m'<< M_D, M_\Sigma$, which is a reasonable assumption to satisfy the oblique parameters}~\cite{Peskin:1990zt, Peskin:1991sw}. 
Then the neutral, singly-, doubly mass eigenvalues and eigenstates are respectively given by~\cite{Nomura:2017ezy}
\begin{align} 
&D_N\approx{\rm diag}(M_D-(2mm')^{1/2}, M_D+(2mm')^{1/2}, M_\Sigma) ,\
 V\approx 
\left[\begin{array}{ccc}
-\frac{i}{\sqrt2} & \frac{i}{\sqrt2}  & i(-\delta'+\delta) \\ 
\frac{1}{\sqrt2} & \frac{1}{\sqrt2}  & \delta'+\delta \\
-\frac{\delta}2  &-\frac{\delta'}2 & 1 \\ 
\end{array}\right],
\\
&D_\pm \approx{\rm diag} (M_D-\frac12\sqrt{3m^2+m'^2}, M_D+\frac12\sqrt{3m^2+m'^2}, M_\Sigma),\quad
 V_C\approx \left[\begin{array}{ccc}
1& 0& \frac{\delta'}{2} \\ 
0& 1 & -\frac{\sqrt3}{2}\delta \\
\frac{\delta'}{2}  &  -\frac{\sqrt3}{2}\delta &1 \\ 
\end{array}\right],\\
&D_{\pm\pm}\approx{\rm diag}(M_D, M_\Sigma),\quad
 V_{CC}\approx 
\left[\begin{array}{cc}
1  & -\delta' \\ 
{\delta} &1 \\
\end{array}\right],
\end{align}
where we consider one generation, $\delta^{(')}\equiv m^{(')}/(M_\Sigma-M_D)$; ($m^{(')}<<M_\Sigma-M_D$), and the singly-charged sector does not have nonzero perturbations at the first order,

\section{Numerical analyses}
Her we have  numerical analysis in a benchmark point.
First of all, we fix the following parameters:
\begin{align}
& (m_{1},m_2,m_3)=(1,5,10)\ {\rm GeV},\quad (m'_{1},m'_2,m'_3)=(2,10,50)\ {\rm GeV},
\nn\\ &(M_{D_1},M_{D_2},M_{D_3})=(M_{\Sigma_1}+500,M_{\Sigma_2}+500,M_{\Sigma_3}+500)\ {\rm GeV}.
\end{align}
{Note that here we adopt $m_{1-3}$ and $m'_{1-3}$ to be around 1 GeV to 50 GeV that corresponds to $g(g') \sim \mathcal{O}(0.01)-\mathcal{O}(0.1)$ since $m(m') = g(g') v/\sqrt{2}$ as we discussed above.
Thus we chose the magnitude of $m(m')$ so that the corresponding Yukawa couplings are not very small. }
Also we assume all the components of $H_5$  to be degenerate.
Here we consider the case of fermionc DM which is $\Sigma_R^0$ in the first generation.~\footnote{Since a bosonic DM candidate of $\phi^0_{5_R}$ dominant can decay into 4 SM Higgs via 5 dimensional operator $\lambda H_2^\dag H_2 H_2^\dag H_2 H_5$, it could be shorter than the age of Universe even though cut-off scale is Planck scale if $\lambda={\cal O}$(1)}.
Then we have just refer to the detailed results in ref.~\cite{Cirelli:2005uq, Cirelli:2007xd} for relic density of $\Sigma_R^0$. 
Note here that since the quintet Majorana fermions $\Sigma_R$ does not have Yukawa interactions with SM particles at tree level, the result is basically the same as the case of minimal quintet fermion~\cite{Cirelli:2005uq, Cirelli:2007xd}. Also as we assumed above, lepton number is conserved at energy scale higher than cut-off 
so that non-renormalizable operator inducing decay of $\Sigma_R$ {such as ($[\bar\psi_LH^*_2\Sigma_R]H^\dag_2 H_2$)} does not appear, guaranteeing stability of DM~\cite{Cai:2016jrl,Sierra:2016qfa}.
Here we neglect the mixing between the neutral components of $\psi$ and $\Sigma$ that is a natural assumption to satisfy the constraint of oblique parameters. Even though there is the mixing, the decaying process is induced via 8 mass dimension
which can be sufficiently tiny.

\begin{figure}[tb]\begin{center}
\includegraphics[width=8cm]{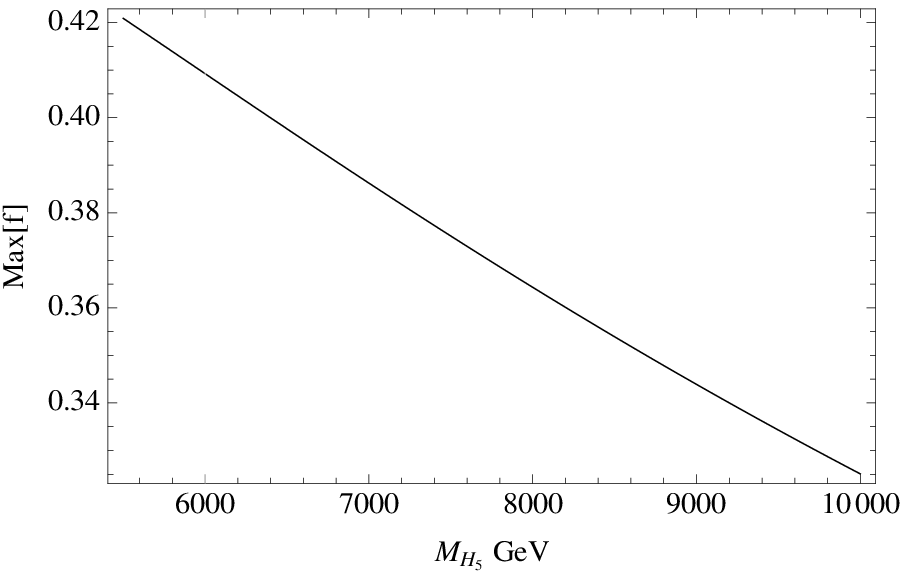}
\includegraphics[width=8cm]{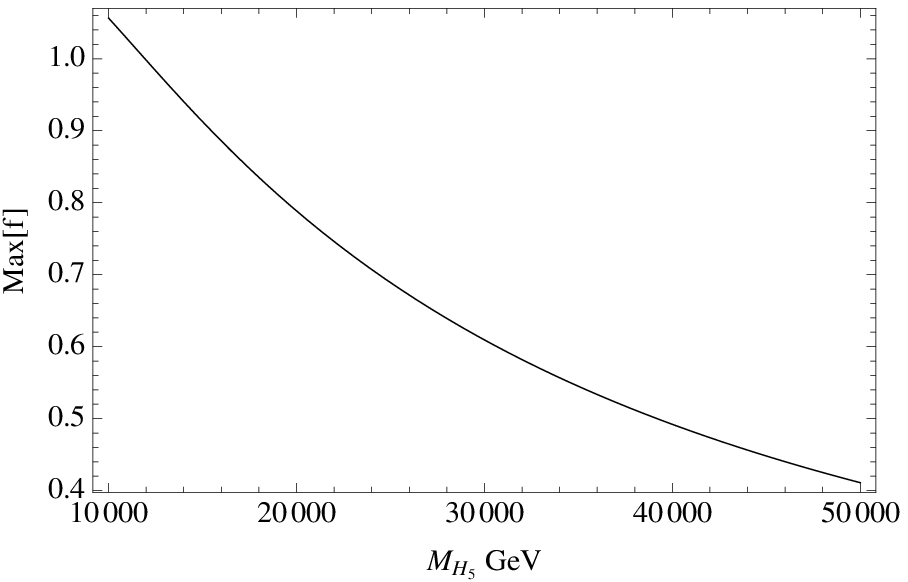}
\caption{A line of the maximum absolute value of $f$(=Max[f]) in terms of  $M_{H_5}$, where Max[$f$]
is 0.41(1.0) in the left(right) side that  satisfies the perturbation limit ($\sim\sqrt{4\pi}$). }   \label{fig:f}
\end{center}\end{figure}

\begin{figure}[tb]\begin{center}
\includegraphics[width=8cm]{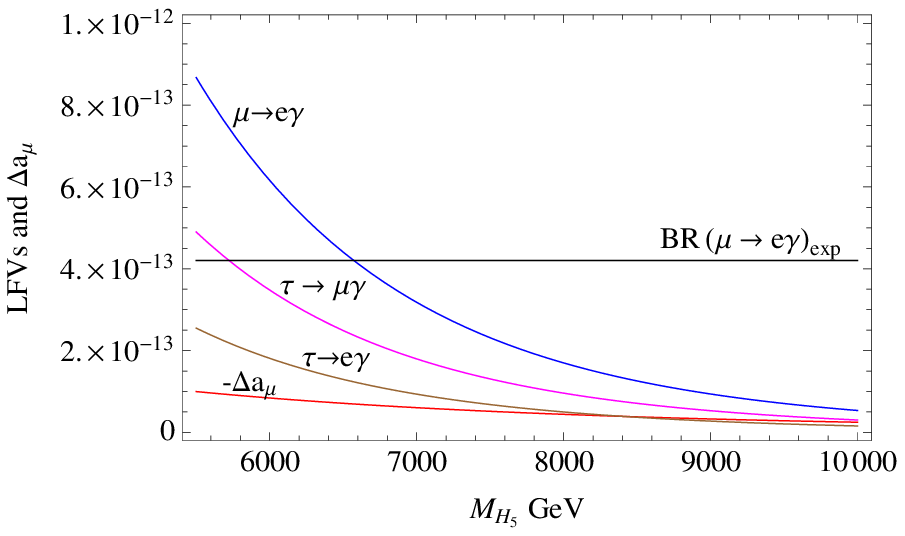}
\includegraphics[width=8cm]{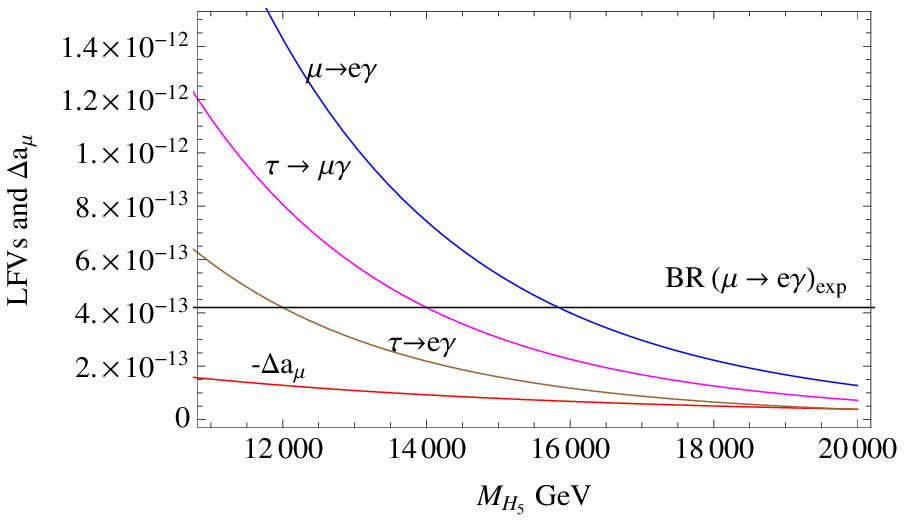}
\caption{Lines of LFV processes and $\Delta a_\mu$ in terms of  $M_{\Sigma_1}$, where
blue, magenta, brown, and red lines respectively denote the theoretical bounds on BR($\mu\to e\gamma$), BR($\tau\to \mu\gamma$),
BR($\tau\to e\gamma$), and -$\Delta a_\mu$, while the black horizontal line is the experimental bound on BR($\mu\to e\gamma$).
The left(right) side corresponds to $M_{\Sigma_1}\sim$4.4(10) TeV. 
}   \label{fig:lfvs}\end{center}\end{figure}

The typical DM mass is $M_{\Sigma_1}\sim$4.4(10) TeV to satisfy the relic density estimated by perturbative(including non-perturbative effect) calculation~\cite{Cirelli:2007xd}.  
Then we fix the other parameters as follows:
\begin{align}
& O _{12,23,13}=\pi+5.7i(6.0i) ,\quad m_I=M_{H_5}+10\ {\rm GeV},\nn\\
& (M_{\Sigma_2},M_{\Sigma_3})=(M_{\Sigma_1}+1000, M_{\Sigma_1}+2000)\ {\rm GeV},
\end{align}
where $O_{12,23,13}$ are arbitral mixing matrix with complex values that are introduced in the neutrino sector~\footnote{In general, there are 21 free parameters. But here we simply reduce them to be three.}, and $M_{H_5}$ is considered as a free parameter.
{Here the values of $O_{12,23,13}$ are chosen so that new physics contributions to muon $g-2$ become as large as possible. 
Also we summarize benchmark values of new fermion masses in Table~\ref{tab:2}. }
Fig.~\ref{fig:f} shows the maximum absolute value of $f$(=Max[f]) in terms of  $M_{H_5}$, which suggests Max[$f$]
is 0.41(1.0) for $M_{\Sigma_1}\sim$4.4 TeV(left side) and 10 TeV(right side) that satisfies the perturbation limit ($\sim\sqrt{4\pi}$) very well.
On the other hand, Fig.~\ref{fig:lfvs} represents various LFV processes and $\Delta a_\mu$ in terms of  $M_{\Sigma_1}$ for $M_{\Sigma_1}\sim$4.4 TeV(left side) and 10 TeV(right side). 
Here blue, magenta, brown, and red lines respectively denote the theoretical bounds on BR($\mu\to e\gamma$), BR($\tau\to \mu\gamma$),
BR($\tau\to e\gamma$), and $-\Delta a_\mu$. The figure suggests 6.6(16) TeV$\le M_{H_5}$ that comes from the bound on BR($\mu\to e\gamma$).
{Notice that our muon $g-2$ is negatively induced as $-\Delta a_\mu\sim {\cal O}(10^{-15})$ when we assume degenerate masses of $H_5$ components; 
the terms giving negative contribution in Eq.~(\ref{eq:aR}) dominate the terms inducing positive contribution and this situation could be modified considering hierarchy among masses of $H_5$ components.
The predicted value of $\Delta a_\mu$ is negative and its absolute value is very small compared with the discrepancy between experimental data and the SM prediction, $\Delta a_\mu \sim 10^{-9}$. 
Thus the contributions in our new particles are not suitable to explain the discrepancy.  }

\begin{table}[t!]
\begin{tabular}{|c||c|c|c|c|c|c|}\hline\hline  
Particle & $\Psi^{Q}_1$ & $\Psi^{Q}_2$ & $\Psi^{Q}_3$ & $\Psi^{Q}_4$ & $\Psi^{Q}_5$ & $\Psi^{Q}_6$  \\ \hline
mass [TeV] & 4.4(10.0) & 5.4(11.0) & 6.4(12.0) & 4.9(10.5) & 5.9(11.5) & 6.9(12.5)   \\ \hline
\end{tabular}
\caption{The benchmark mass spectrum for new fermions in the model where $Q= \{\pm \pm, \pm, 0\}$. }\label{tab:2}
\end{table}

\begin{figure}[tb]\begin{center}
\includegraphics[width=8cm]{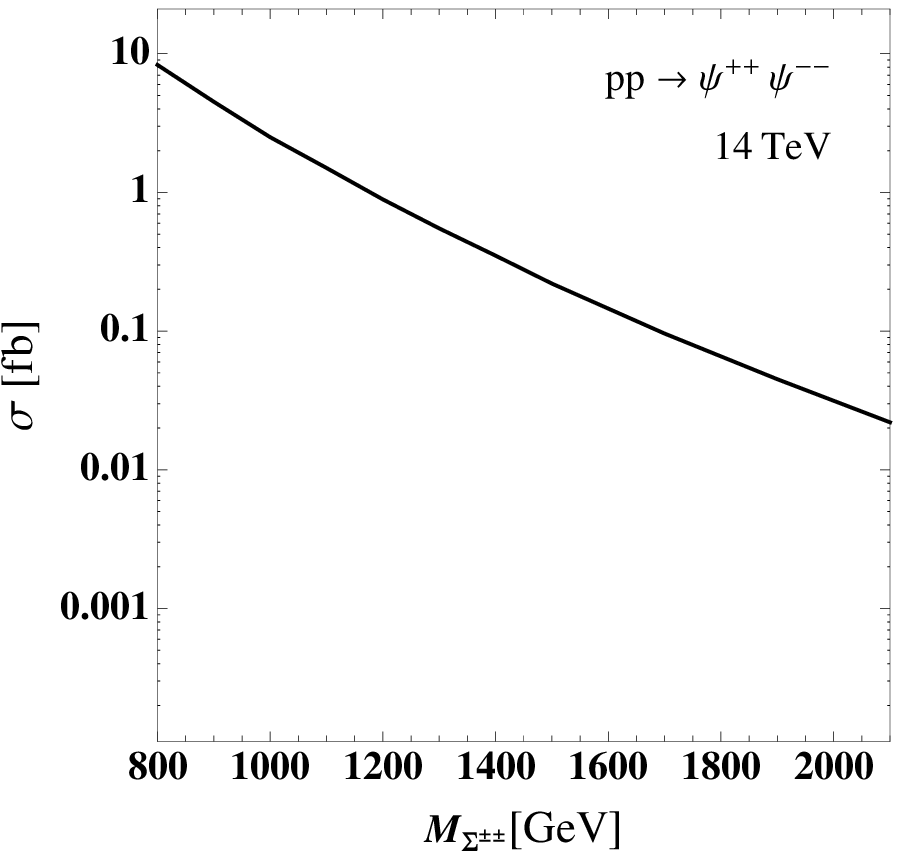}
\includegraphics[width=8cm]{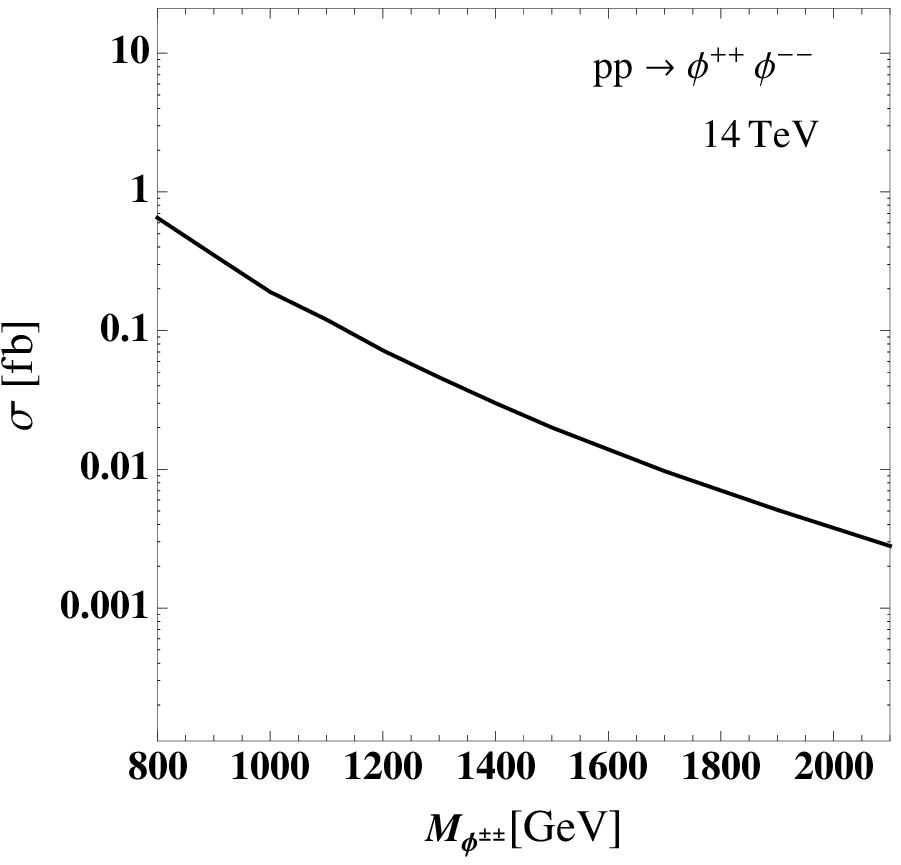}
\caption{Doubly charged fermion and scalar production cross sections as functions of their masses at the LHC 14 TeV including both Drell-Yan and photon fusion processes.
}   \label{fig:CX}\end{center}\end{figure}

Here let us briefly comments possible collider physics of our model.
Rich phenomenology at collider experiments can be induced since there are many exotic charged particles from large $SU(2)$ multiplet scalars and fermions as we show in Eq.~(\ref{eq:H5})-(\ref{eq:sigmaR}).
The charged particles in the multiplets can be produced through electroweak interactions at hadron collider experiments.
For $\mathcal{O}(1)$ TeV mass scale, production cross sections can be $\mathcal{O}(1)$ fb scale for doubly charged fermion at the LHC 13 TeV~\cite{Nomura:2017abu} 
and sizable number of events will be obtained with integrated luminosity of $\mathcal{O}(100)$ fb$^{-1}$.
{We also show doubly charged fermion and scalar production cross section as functions of their mass in Fig.~\ref{fig:CX} where we included both Drell-Yan process and 
photon fusion process $\gamma \gamma \to \Sigma^{++} \Sigma^{--} (\phi^{++}_5 \phi^{--}_5)$, and the center of mass energy $\sqrt{s} = 14$ TeV is applied.}
Then produced charged particles decay into lighter component in these large multiplets with SM gauge bosons $W^\pm/Z$ which will be off-shell state as the mass difference is smaller than $W^\pm/Z$ boson mass.
The signatures of our new particles could be obtained as multi-particle states including charged leptons, jets and missing transverse momentum.
Note, however, that in our benchmark point the production cross section will be too small. 
In such a case, collider experiment with higher energy is required such as 100 TeV hadron collider.
Since the signal is complicated, detailed analysis is beyond the scope of this paper~\footnote{Collider phenomenology of charged particles in large multiplet is discussed in, for example, refs.~\cite{delAguila:2013yaa,delAguila:2013mia, Nomura:2017abu,Chala:2018ari}.}. 
{The mass scale can be $\mathcal{O}(1)$ TeV if we do not require neutral component of the multiplet to satisfy observed relic density of DM, 
and we would get detectable number of signal event at the LHC 14 TeV in near future. Note that in lower mass scale the relic density of the neutral component becomes much smaller than observed relic density 
and it is not excluded by cosmological constraints.  
In addition we expect displaced vertex signature since lifetime of charged component could be long as $\tau \gtrsim $ cm; 
for $M_\Sigma \sim 4.4$ TeV, it is estimated to be $\tau \sim 1.8$ cm~\cite{Cirelli:2005uq} and the lifetime will be longer for lighter mass scale.
We left further analysis in future work.
}

\section{Summary and Conclusions}
We have constructed a model in which neutrino mass is induced at one-loop level introducing fields with large multiplet of $SU(2)_L$. 
Thanks to large representations, our model has been realized without any additional symmetries, and we have formulated lepton and new fermion sector such as mass matrices, LFVs, and muon $g-2$.
Furthermore, our model is valid up to 100 TeV at most via RGEs, which could be tested by current experiments such as collider.
Due to small mixings among neutral fermions that are required by the oblique parameters and heavier extra masses that come from lower mass bound on DM mass 4.4(10) TeV estimated by perturbative(including non-perturbative effect) calculation, our LFV processes are enough suppressed but muon $g-2$ is also suppressed.  
 Finally, we have brief commented possibility to detect at LHC or 100 TeV future collider.

\section*{Acknowledgments}
This research was supported by an appointment to the JRG Program at the APCTP through the Science and Technology Promotion Fund and Lottery Fund of the Korean Government. This was also supported by the Korean Local Governments - Gyeongsangbuk-do Province and Pohang City (H.O.)
H. O. is sincerely grateful for KIAS and all the members.

\end{document}